\documentclass[%
 reprint,
 amsmath,amssymb,
 aps,
]{revtex4-2}

\usepackage{graphicx}% Include figure files
\usepackage{dcolumn}% Align table columns on decimal point
\usepackage{bm}% bold math
\usepackage{subfig}
\usepackage{hyperref}% add hypertext capabilities
\usepackage{flafter}
\usepackage{placeins}
\usepackage{xcolor}
\usepackage{booktabs}   % For \toprule, \midrule, \bottomrule
\usepackage{amsmath}    % For math symbols
\begin{document}

\preprint{APS/123-QED}

\title{Probing Criticality Using GMM-Based Potentials}% Force line breaks with \\

\author{Shashank Sharma}
\email{ssharma23@iitk.ac.in}
\affiliation{%
 Department of Physics, Indian Institute of Technology Kanpur, India
}

\author{Dipankar Chakrabarti}
\email{dipankar@iitk.ac.in}
\affiliation{%
 Department of Physics, Indian Institute of Technology Kanpur, India
}

\author{Vipul Arora}
\email{vipul.arora@kuleuven.be}
\affiliation{%
 Department of Electrical Engineering (ESAT-PSI), KU Leuven, Belgium
}

\date{\today}% It is always \today, today,

\begin{abstract}
% Searching for critical points and measuring the critical exponents of any interesting theory is computationally expensive. We propose a computationally cheaper method for scalar theories to obtain critical points with the same critical exponents using Gaussian Mixture Models (GMMs) that can be sampled with low computational cost.

% We exploit the fact that multiple theories can share the same critical point and construct a distribution using GMMs that will have the same critical point as the theory of interest and can be sampled exactly.\\

Spin models with a given symmetry are easier to sample than scalar theories with the same symmetry on a lattice, as the constrained nature of spin variables enables cheap heat-bath updates. However, this constraint suppresses radial fluctuations, and consequently, spin models cannot be used to study the phenomena of spontaneous symmetry breaking, such as the Higgs phenomenon. To address this, we propose a class of scalar potentials based on Gaussian Mixture Models (GMMs) that are as easy to sample as spin models with a given symmetry. These potentials can be designed to belong to the same universality class as the theory of interest, thereby reproducing its critical properties while enabling efficient sampling. We construct such models for global $\mathbb{Z}_2$ symmetry, $U(1)$ gauge symmetry, and disordered systems. We also verify by numerical experiment that the case of $\mathbb{Z}_2$ symmetry in two dimensions lies in the two-dimensional Ising universality class.

\end{abstract}

\maketitle
\section{Introduction}
Tuning a lattice-regularized quantum field theory to a critical point is central to extracting continuum physics from a discrete formulation. The locality, symmetry, and vacuum degeneracy of the theory determine the location and nature of its critical points. Under the renormalization group flow, critical points flow to scale-invariant fixed points~\cite{Cardy1996}. Critical phase transitions in lattice-regularized systems provide a powerful non-perturbative tool for studying a wide range of physical phenomena, including spontaneous symmetry breaking, the Higgs phenomena~\cite{Bonati2021, Wellegehausen2011, Bonati2021b, Csikor1994, Kajantie2004, Sexty2005}, conformal invariance in systems with disorder~\cite{Chatelain1998, Bagamery2005, Antunes2026} and defects \cite{ Billo2013, Sinha2026}, phase transitions in neural networks\cite{Raya2023, Tamai2025}, etc. At these critical points, microscopic details become irrelevant and universal behavior emerges. Consequently, multiple quantum field theories can belong to the same universality class~\cite{Kadanoff1990}, as exemplified by the equivalence between the Ising model and the real scalar $\phi^4$ theory. 

For tuning a lattice-regularized system to criticality, we need to sample the system across the parameter space, and based on the universality class, we can choose to either work with constrained spin variables or with unconstrained scalar variables. Given an internal symmetry, whether discrete or continuous, spin models with constrained variables are easier to sample than scalar field models on a lattice. The local heat-bath~\cite{Kennedy1985} updates for constrained spin variables are cheaper than those for unconstrained
scalars because the spin variables $s$ are constrained by $|s| = 1$, which suppresses radial fluctuations. However, constrained variables cannot describe a Lorentz-invariant theory and hence a scalar field theory, and, when coupled to gauge fields, they cannot be used to study phenomena of spontaneous symmetry breaking like Higgs phenomena. Also, constrained discrete variables lack well-defined derivatives, so they can not be used for Gradient Flow studies on a lattice~\cite{Sonoda:2019, Carosso:2018, Sonoda:2020, Makino:2018, Morikawa:2024, Sharma2025, Luscher2013}. Therefore, it is necessary to define scalar field variables $\phi$ that possess radial fluctuations, i.e., $|\phi| = \rho$. Sampling such variables requires drawing $\rho$ from a radial distribution $\rho \sim p(\rho)$,  which increases the computational cost.

Given the required properties of a theory such as locality, symmetry, vacuum degeneracy, and the presence of radial fluctuations, one can define an easy-to-sample distribution $p(\rho)$. Moreover, when coupled to gauge fields, this distribution must remain easy to sample. 

% The locality, symmetry, and vacuum degeneracy determine the critical point of the theory.
% The critical points of a quantum field theory (QFT) flow to scale-invariant points under the renormalization group (RG) flow. At these critical points, microscopic details become irrelevant, and universal behavior emerges. Consequently, multiple QFTs can belong to the same universality class \cite{Kadanoff1990}. A well-known example is the equivalence between the Ising model and the real scalar $\phi^4$ theory.

Leveraging this property, we propose to design a distribution using a Gaussian Mixture Model (GMM) \cite{Fruhwirth2019}.  GMMs are used to estimate priors for low-dimensional multivariate non-Gaussian distributions and have found applications in lattice field theory, including multilevel generative sampling \cite{Singha2025ICLR, Singha2026}, as well as in the analysis of spontaneous symmetry breaking in generative diffusion models \cite{Raya2023}. By wrapping spin variables with Gaussian noise, we define $\rho \sim \mathcal{N}(0, \sigma)$, such that the resulting distribution belongs to the same universality class as the quantum field theory (QFT) of interest. These GMM-based distributions are as easy to sample as spin models, since the radial distribution is Gaussian noise, which is very cheap to sample alongside a local heat-bath update. We will show later that even when adding linear or quadratic disorder and coupling to gauge fields, the local radial distribution remains Gaussian with an effective mean and an effective variance. Therefore, rather than tuning the original QFT to criticality, which is computationally demanding, we can tune a GMM-based distribution to criticality.  Since the two systems share the same underlying conformal field theory (CFT) at criticality, the critical exponents can be measured reliably.

We demonstrate the utility of this approach for several lattice field theories, including the two-dimensional scalar theory with $\mathbb{Z}_2$ symmetry, disordered systems, defect systems, and a complex scalar coupled to a $ U(1)$ gauge field.

\section{Gaussian Mixture Models (GMMs)}
 
A GMM is a probabilistic model that represents a distribution as a weighted sum of multiple Gaussian components. Formally, the likelihood of a data point $\mathbf{x} \in \mathbb{R}^D$ is given by
\begin{equation}
p(\mathbf{x}) = \sum_{k=1}^{K} \pi_k \, \mathcal{N}(\mathbf{x} \mid \boldsymbol{\mu}_k, \boldsymbol{\Sigma}_k),
\end{equation}
where $K$ is the number of components, $\pi_k$ are the mixing coefficients satisfying $\sum_{k} \pi_k = 1$ and $\pi_k \ge 0$, and $\mathcal{N}(\mathbf{x} \mid \boldsymbol{\mu}_k, \boldsymbol{\Sigma}_k)$ denotes a Gaussian distribution with mean $\boldsymbol{\mu}_k$ and covariance $\boldsymbol{\Sigma}_k$. GMMs are widely used for density estimation, clustering, and as flexible approximators of complex multimodal distributions \cite{Ruzgas2025, Lu2025, Thrun2015, McNicholas2016, Lu2021, Zhao2025}. 

\section{GMM-Based $\mathbb{Z}_2$-symmetric potential for real scalar}
\label{sec:Z2_GMM}

We begin with the standard Euclidean lattice action for a scalar field in d-dimensions
\begin{align}
S[\phi] = \sum_{x} \Big[-\phi_x
\frac{1}{2} \sum_{\mu=1}^{d} ( \phi_{x+\hat{\mu}}+\phi_{x-\hat{\mu}}-2\phi_x) + V(\phi_x)
\Big],
\end{align}
where $\phi_x$ is a real-valued scalar defined on each lattice site. We define a local $\mathbb{Z}_2$-symmetric GMM-Based potential,
\begin{align}
e^{-V(\phi_x)} 
&= \frac{1}{2} \exp\left(-\frac{(\phi_x - v)^2}{2\sigma^2}\right)
\nonumber \\
&\quad + \frac{1}{2} \exp\left(-\frac{(\phi_x + v)^2}{2\sigma^2}\right),
\label{eq:Z2_GMM_potential}
\end{align}
with mean $v$ and variance $\sigma^2$. The compact form of the lattice action would look like
\begin{align}
S[\phi] = \sum_{x} \Big[-\phi_x
\frac{1}{2} \sum_{\mu=1}^{d} ( \phi_{x+\hat{\mu}}+\phi_{x-\hat{\mu}}-2\phi_x) +\frac{\phi_x^2 + v^2}{2\sigma^2} 
\nonumber \\
- \mathrm{log~cosh}\left(\frac{\phi_x v}{\sigma^2}\right)
\Big].
\label{eq:Z2_GMM_action}
\end{align}
Thus, the tunable bare parameters are $v$ and $\sigma^2$, with \(v,\sigma>0\). Notice that the potential is analytic for all values of $\phi_x$.

Introducing an auxiliary discrete variable $z_x = \pm 1$ , Eq.~\eqref{eq:Z2_GMM_potential} can be written as
$e^{-V(\phi_x)}
= \sum_{z_x = \pm 1}
\frac{1}{2}
\exp\left(-\frac{(\phi_x - z_x v)^2}{2\sigma^2}\right)$, the action for the joint distribution is
\begin{align}
S[\phi, z] = \sum_x \Big[-\phi_x
\frac{1}{2} \sum_{\mu=1}^{d} ( \phi_{x+\hat{\mu}}+\phi_{x-\hat{\mu}}-2\phi_x)
\nonumber\\+ \frac{1}{2\sigma^2} (\phi_x - v z_x)^2
\Big],
\label{eq:Z2_GMM_joint_action}
\end{align}
This action will give the same partition function as the marginal action over $\phi$ only. A similar potential as Eq~\eqref{eq:Z2_GMM_action} appears when Hubbard-Stratonovich transformations \cite{Hubbard, Stratonovich} are applied to the Ising model~\cite{Ostmeyer2021} and the infinite-range Ising model \cite{Semenoff}.
% Expanding the quadratic term,
% \begin{align}
% \frac{(\phi_x - z_x v)^2}{2\sigma^2}
% = \frac{\phi_x^2}{2\sigma^2}
% - \frac{z_x v}{\sigma^2}\phi_x
% + \frac{v^2}{2\sigma^2},
% \end{align}
\subsection{Heat-Bath Sampling}
\label{subsec:Heat_Bath_Updates}
To derive the local update rule, we isolate terms depending on a single site in Eq.~\eqref{eq:Z2_GMM_joint_action}, define the nearest-neighbor sum $s_x = \frac{1}{2} \sum_{y \in nn(x)} \phi_y$, and rewrite
\begin{align}
S_x(\phi_x, s_x,z_x) =
\left(d + \frac{1}{2\sigma^2}\right)\phi_x^2
- \frac{z_x v}{\sigma^2}\phi_x
- s_x \phi_x.
\end{align}% \begin{align}
% S_x =
% \left(2 + \frac{1}{2\sigma^2}\right)\phi_x^2
% - \frac{z_x v}{\sigma^2}\phi_x
% - \frac{1}{2} \sum_{y \in nn(x)} \phi_x \phi_y.
% \end{align}
Keeping $s_x$ and $z_x$ fixed, the conditional probability becomes
\begin{align}
P(\phi_x \mid z_x, s_x)
\propto
\exp\Big[
&- \left(d + \frac{1}{2\sigma^2}\right)\phi_x^2
\nonumber \\
&+ \left(\frac{z_x v}{\sigma^2} + s_x\right)\phi_x
\Big].
\end{align}
 
Define $A = d + \frac{1}{2\sigma^2}$ and $
B = \frac{z_x v}{\sigma^2} + s_x$, and completing the square, we obtain

% Then
% \begin{align}
% P(\phi_x \mid z_x, \text{neighbors}) \propto \exp\left[-A \phi_x^2 + B \phi_x\right].
% \end{align}

% Completing the square,
% \begin{align}
% - A \phi_x^2 + B \phi_x
% = -A \left(\phi_x - \frac{B}{2A}\right)^2
% + \frac{B^2}{4A},
% \end{align}
%
%we obtain
\begin{align}
P(\phi_x \mid z_x, s_x)
\propto
\exp\Big[
- A \left(\phi_x - \frac{B}{2A}\right)^2 
\Big].
\end{align}
Summing over $z_x$,
\begin{align}
P(\phi_x \mid s_x)
= w_+ \;
\mathcal{N}(
\phi_x |\mu_x^{(+)},
\;
\sigma_\text{eff}
)
\nonumber \\
+\; w_- \;
\mathcal{N}(
\phi_x |\mu_x^{(-)},
\;
\sigma_\text{eff}
),
\label{eq:heat_bath}
\end{align}

where
\begin{align}
w_+ = \frac{1}{2}\left[1 + \tanh\left(\frac{v s_x}{2A \sigma^2}\right)\right], \quad
w_- = 1 - w_+
%\frac{1}{2}\left[1 - \tanh\left(\frac{v s_x}{A \sigma^2}\right)\right],
\end{align}
or equivalently $w_+ = \sigma\!\left(\frac{v s_x}{A \sigma^2}\right),
w_- = \sigma\!\left(-\frac{v s_x}{A \sigma^2}\right)$, where $ \sigma(t) = \frac{1}{1 + e^{-t}}$. The effective means and the effective variance are
\begin{align}
\mu_x^{(\pm)} =
\frac{\pm\frac{v}{\sigma^2} + s_x}
{2\left(d + \frac{1}{2\sigma^2}\right)},
% \mu_x^{(-)} =
% \frac{-\frac{v}{\sigma^2} + s_x}
% {2\left(2 + \frac{1}{2\sigma^2}\right)},
\quad
\sigma_{\text{eff}}^2 =
\frac{1}{2\left(d + \frac{1}{2\sigma^2}\right)}.
\end{align}
Thus, the local update rule given by Eq.~\eqref{eq:heat_bath} corresponds exactly to sampling from a Gaussian mixture whose means are shifted by the neighboring field $s_x$, while the kinetic term modifies the variance. Eq.~\eqref{eq:heat_bath} allows rejection-free, low-cost local heat-bath updates~\cite{Kennedy1985, Faraz2023}.

\subsection{Sampling Using Local Hamiltonian Dynamics}
\label{sec:local_hamiltonian_dynamics}

For fixed auxiliary variables $\{z_x\}$, the joint action given by Eq~\eqref{eq:Z2_GMM_joint_action} is
quadratic in $\phi$; this property allows for efficient update rules other than the heat-bath updates, which we discussed in the previous sec.~\ref{subsec:Heat_Bath_Updates}. As the sites are conditionally independent given the neighboring sites, we can also perform local updates using Hamiltonian dynamics~\cite{Horvath1998}. Again, the conditional distribution of a single site variable $\phi_x$ is a Gaussian
\begin{equation}
P(\phi_x|z_x,s_x)\propto\exp\!\Big[ -A\phi_x^2 + B\phi_x\Big],
\label{eq:cond}
\end{equation}
Completing the square gives mean \(\mu_x=\frac{B}{2A}\) and variance \(\frac{1}{2A}\). 
To sample the conditional given by Eq.~\eqref{eq:cond} exactly, we introduce a momentum
$p_x\sim\mathcal{N}(0,1)$ and the Hamiltonian
$H=\tfrac12 p_x^2 + A\phi_x^2 - B\phi_x$.  Hamilton's equations
$\dot\phi_x=p_x$, $\dot p_x=-2A\phi_x+B$, describe a harmonic
oscillator of frequency $\omega=\sqrt{2A}$.  They can be integrated
analytically:
\begin{equation}
\begin{aligned}
\phi_x(\tau) &= \phi_x(0)\cos\omega\tau + \frac{p_x(0)}{\omega}\sin\omega\tau
               + \mu_x(1-\cos\omega\tau),\\
p_x(\tau) &= p_x(0)\cos\omega\tau - \omega\phi_x(0)\sin\omega\tau
            + \frac{B}{\omega}\sin\omega\tau.
\end{aligned}
\label{eq:eom}
\end{equation}
The trajectory of length $\tau$ is exact; no Metropolis test is required.
The special choice $\tau=\pi/\omega$ reduces Eq~\eqref{eq:eom} to a simple
reflection about the mean,
\begin{align}
    \phi_x \to -\phi_x+2\mu_x,~p_x\to -p_x.
\end{align}
Which, together with the
momentum refreshment generates a valid update.

For $\tau = \pi/(2\omega)$, Eq.~\eqref{eq:eom} reduces to
\begin{align}
\phi_x(\tau) = \mu_x + \frac{p_x(0)}{\omega},
\qquad
p_x(\tau) = -\omega\,\phi_x(0) + \frac{B}{\omega}.
\end{align}
Since $p_x(0) \sim \mathcal{N}(0,1)$, the new field value is distributed as
\begin{align}
\phi_x(\tau) \sim \mathcal{N}\!\left(\mu_x, \frac{1}{2A}\right),
\end{align}
which is exactly the conditional distribution $P(\phi_x \mid z_x, s_x)$ given by Eq.~\eqref{eq:cond}, reproducing heat-bath updates discussed in Sec.~\ref{subsec:Heat_Bath_Updates}.
After updating all $\phi_x$, the auxiliary
variables $z_x$ are updated by an exact Gibbs step,
\begin{equation}
P(z_x=\pm1\mid\phi_x) \propto \exp\!\Big(\pm\frac{v}{\sigma^2}\phi_x\Big).
\label{eq:gibbs_z}
\end{equation}
The algorithm is exact because the exact
Gaussian conditional given by Eq.~\eqref{eq:cond} is sampled without approximation. In this work, for numerical experiments, we have used heat-bath updates only.

\subsection{Integrating out {$\phi$}}

We begin by taking the continuum limit of the lattice action given by Eq~\eqref{eq:Z2_GMM_joint_action}. 
With lattice spacing $a$, the fields $\phi_x$ and \(z_x\) are now defined on continuous space $\phi(x)$ and \(z(x)\), respectively, where \(x \in \mathbb{R}^d\), and the sum over lattice sites becomes an integral,
$\sum_x \to a^{-d} \int d^d x$. The discrete Laplacian becomes 
the continuum Laplacian, and we define the continuum fields with appropriate scaling so that the action takes the form
\begin{equation}
S[\phi,z] = \int d^d x \Big[-\frac{1}{2}\phi\nabla^2 \phi 
+ \frac{1}{2\sigma^2}(\phi - v z)^2\Big],
\label{eq:continuum_action}
\end{equation}
where $z(x) \in \{\pm 1\}$ is the continuum auxiliary field. 
Given \(z\), this action is quadratic in $\phi$, which allows us to integrate 
it out exactly.

Expanding the potential term in Eq.~\eqref{eq:continuum_action} gives
\begin{equation}
S[\phi,z] = \int d^d x \Big[-\frac{1}{2}\phi\nabla^2 \phi 
+ \frac{1}{2\sigma^2}\phi^2 - \frac{v}{\sigma^2}\phi z 
+ \frac{v^2}{2\sigma^2}z^2\Big].
\end{equation}
Since $z^2 = 1$ for $z(x) = \pm 1$, the last term is just a 
constant that can be absorbed into the normalization of the 
partition function. The action thus becomes
\begin{equation}
S[\phi,z] = \int d^d x \Big[~\frac{1}{2}\phi\left(-\nabla^2 
+ \frac{1}{\sigma^2}\right)\phi - \frac{v}{\sigma^2}\phi z \Big],
\label{eq:quadratic_action}
\end{equation}
up to an irrelevant additive constant.

To integrate out $\phi$, we complete the square in Eq.~\eqref{eq:quadratic_action}. 
Define the operator $\mathcal{A} = -\nabla^2 + 1/\sigma^2$, 
with Green's function $G(x-y)$ satisfying
\begin{equation}
\mathcal{A} G(x-y) = \delta^{(d)}(x-y).
\end{equation}
Then the action can be rewritten as
\begin{equation}
S[\phi,z] = \frac{1}{2}\int d^d x \, \phi(x) \mathcal{A} \phi(x)
- \frac{v}{\sigma^2}\int d^d x \, \phi(x) z(x).
\end{equation}
The functional integral gives the partition function
\begin{equation}
Z = \sum_{\{z(x)\}} \int \mathcal{D}\phi \, e^{-S[\phi,z]}.
\end{equation}
Performing the Gaussian integral over $\phi$, the partition function 
thus reduces to
\begin{equation}
Z =  {N} \sum_{\{z(x)\}} 
\exp\Big[~\frac{v^2}{2\sigma^4}\int d^d x \int d^d y \, 
z(x) G(x-y) z(y)\Big],
\end{equation}
where ${N}$ is an overall normalization constant. 
Consequently, the effective action for the $z$ field is
\begin{equation}
S_{\text{eff}}[z] = -\frac{v^2}{2\sigma^4}
\int d^d x \int d^d y \, z(x) G(x-y) z(y),
\label{eq:Seff_position}
\end{equation}

The effective action in Eq. \eqref{eq:Seff_position} describes 
a long-range Ising model, where the spins $z(x) \in \{\pm 1\}$ 
interact via the nonlocal kernel $G(x-y)$. For $\sigma \to 0$, the kernel becomes local, and the model 
reduces to the standard short-range Ising model.

\subsection{Critical Point Condition}

The GMM $\mathbb{Z}_2$-symmetric potential reads
\begin{align}
V(\phi) = \frac{\phi^2 + v^2}{2\sigma^2} 
- \log\cosh\left(\frac{\phi v}{\sigma^2}\right).
\end{align}
The first derivative is
\begin{align}
V'(\phi) = \frac{\phi}{\sigma^2} 
- \frac{v}{\sigma^2} \tanh\left(\frac{\phi v}{\sigma^2}\right),
\end{align}
which vanishes at $\phi = 0$ for all values of $v$ and $\sigma$. The curvature at the origin follows from the second derivative:
\begin{align}
V''(\phi) = \frac{1}{\sigma^2} 
- \frac{v^2}{\sigma^4} \,\text{sech}^2\left(\frac{\phi v}{\sigma^2}\right).
\end{align}
Evaluating at $\phi = 0$ and using $\text{sech}^2(0) = 1$ yields the exact result
\begin{align}
V''(0) = \frac{1}{\sigma^2} - \frac{v^2}{\sigma^4}
= \frac{1}{\sigma^2}\left(1 - \frac{v^2}{\sigma^2}\right).
\end{align}
The nature of the origin is thus determined solely by the dimensionless ratio $ v^2/\sigma^2 $. For $ v^2 < \sigma^2 $, the origin is a local minimum, that is, the potential is unimodal. For $v^2 > \sigma^2$, the origin is a local maximum, that is, the potential is bimodal. The condition $v^2 = \sigma^2$ gives $V''(0) = 0$, the curvature vanishes. The condition
\begin{align}
{v_c^2 = \sigma^2}
\end{align}
follows directly from the exact curvature at $ \phi = 0 $. When the parameters are understood as renormalized couplings $v_R$ and $\sigma_R$, the condition $v_R^2 = \sigma_R^2$ defines a quantum critical point separating the two phases. For the condition \(v_R^2>\sigma_R^2\), the theory becomes ordered and chooses a particular vacuum \(\phi=\pm v\).

\subsection{Discrete variable limit ($\sigma\to 0$)}

If we take the limit $\sigma\to 0$ in the joint action given by Eq.~\eqref{eq:Z2_GMM_joint_action}, the real scalar field $\phi$ is constrained to be $\pm v$ only. That is the joint action $S[\phi,z]$ will depend on only $z$,
\begin{equation}
    S[z] = - v^2 \Sigma_{<x,y>} \ z_x z_y \ +\ \text{constant}.
    \label{eq:Ising_limit}
\end{equation}
This is an Ising model, with $<x,y>$ representing nearest neighbors $x,y$. For \(d=2\), the critical point should be given by Onsager's solution~\cite{Onsager1944}, $v_c^2= \text{ln}(1+\sqrt{2})/2$, that is $v_c = 0.66384$.

\subsection{Including a linear disorder term}

We now introduce a site-dependent linear disorder field, \( h_x \), which couples linearly to the scalar field:
\begin{align}
S \;\rightarrow\; S - \sum_x h_x \phi_x.
\end{align}
Accordingly, the conditional probability is modified to
\begin{align}
P(\phi_x \mid z_x,s_x )
\propto
\exp\Big[
&- \left(d + \frac{1}{2\sigma^2}\right)\phi_x^2
\nonumber \\
&+ \left(
\frac{z_x v}{\sigma^2}
+ s_x
+ h_x
\right)\phi_x
\Big].
\end{align}
Summing over $z_x$, we will again have a similar distribution as for the case without disorder, with
\begin{align}
w_+ = \sigma\!\left(\frac{v (s_x+h_x)}{A \sigma^2}\right), \qquad
w_- = \sigma\!\left(-\frac{v( s_x+h_x)}{A \sigma^2}\right),
\end{align}
and the effective mean and the effective variance are 
\begin{align}
\mu_x^{(\pm)} =
\frac{\pm\frac{v}{\sigma^2} + s_x + h_x}
{2\left(d + \frac{1}{2\sigma^2}\right)},
% \mu_x^{(-)} =
% \frac{-\frac{v}{\sigma^2} + s_x}
% {2\left(2 + \frac{1}{2\sigma^2}\right)},
\quad
\sigma_{\text{eff}}^2 =
\frac{1}{2\left(d + \frac{1}{2\sigma^2}\right)},
\end{align}
respectively. Similarly, the local conditional distribution in the presence of a quadratic disordered term and a linear or quadratic defect term will be an exact Gaussian distribution.

\section{GMM-Based $\text{U}(1)$-Symmetric Potential for Complex Scalar Coupled With Gauge Field}

We construct a $\text{U}(1)$-symmetric local potential for the complex scalar field $\Phi_x \in \mathbb{C}$ as a continuous mixture of Gaussian distributions centered on a circle of radius $v > 0$. This representation is manifestly gauge-invariant and will allow us to introduce an auxiliary angular field $\theta_x$ that greatly simplifies Monte Carlo updates.
\subsubsection{Local Potential as a Continuous Gaussian Mixture}

Define the local Boltzmann weight via the average over orientations:
\begin{align}
e^{-V(\Phi_x)} = \frac{1}{2\pi} \int_{0}^{2\pi} d\theta_x \; 
\exp\left(-\frac{|\Phi_x - v e^{i\theta_x}|^2}{2\sigma^2}\right),
\end{align}
where $v,\sigma > 0$, \(\sigma\) controls the width of each Gaussian component. 
The integral is evaluated by writing $\Phi_x = \rho_x e^{i\varphi_x}$:
\begin{align}
|\Phi_x - v e^{i\theta_x}|^2 = \rho_x^2 + v^2 - 2v\rho_x\cos(\varphi_x - \theta_x).
\end{align}
Hence,
\begin{align}
e^{-V(\Phi_x)} = \frac{1}{2\pi} e^{-\frac{\rho_x^2+v^2}{2\sigma^2}}
\int_0^{2\pi} d\theta_x \; \exp\left(\frac{v\rho_x}{\sigma^2}\cos(\varphi_x - \theta_x)\right).
\end{align}
The integral yields the modified Bessel function of the first kind $I_0$:
\begin{align}
\int_0^{2\pi} e^{\kappa \cos(\theta-\varphi)} d\theta = 2\pi I_0(\kappa).
\end{align}
Thus the potential becomes
\begin{align}
{e^{-V(\rho_x)} = e^{-\frac{\rho_x^2+v^2}{2\sigma^2}} \; I_0\!\left(\frac{v\rho_x}{\sigma^2}\right)}.
\end{align}
This potential is $\text{U}(1)$-symmetric, as it depends only on $\rho_x = |\Phi_x|$.

\subsubsection{Joint Lattice Action with Auxiliary Angular Field}

Using the integral representation of the potential, we introduce an auxiliary field $\theta_x \in [0,2\pi)$ at each site and write the joint probability distribution as
\begin{align}
P(\Phi,\theta,U) \propto e^{-S[\Phi,\theta,U]},
\end{align}
with the action
\begin{align}
S[\Phi,\theta,U] = S_{\text{kin}}[\Phi,U] + S_{\text{pot}}[\Phi,\theta] + S_{\text{plaq}}[U].
\end{align}
The gauge-covariant kinetic term is
\begin{align}
S_{\text{kin}}[\Phi,U] =\sum_{x}- \frac{1}{2}\Phi_x^\dagger\sum_{\mu=1}^{d}\Bigl(U_{x,\mu}\Phi_{x+\hat{\mu}} + U_{x-\hat{\mu},\mu}^\dagger\Phi_{x-\hat{\mu}}
\nonumber\\
-2\Phi_x\Bigr),
\end{align}
where $U_{x,\mu} \in \text{U}(1)$ are the link variables. 
The potential term arising from the Gaussian mixture is
\begin{align}
S_{\text{pot}}[\Phi,\theta] = \frac{1}{2\sigma^2}\sum_x \bigl|\Phi_x - v e^{i\theta_x}\bigr|^2.
\end{align}
The pure gauge action (plaquette term) is the standard Wilson form \cite{Rothe2012}
\begin{align}
S_{\text{plaq}}[U] = \beta\sum_{x}\sum_{\mu<\nu}
\Big
[1 - \operatorname{Re}\bigl(U_{x,\mu}U_{x+\hat{\mu},\nu}U_{x+\hat{\nu},\mu}^\dagger U_{x,\nu}^\dagger\bigr)\Big].
\end{align}
The action is invariant under the local \(\text{U}(1)\) gauge transformations
\begin{align}
\Phi_x \to e^{i\alpha_x}\Phi_x,
U_{x,\mu} \to e^{i\alpha_x}U_{x,\mu}e^{-i\alpha_{x+\hat{\mu}}},
\theta_x \to \theta_x + \alpha_x.
\end{align}

\subsubsection{Local Conditional Distribution in the Presence of a Gauge Field}

We now derive the conditional probabilities for updating a single site $x$ while keeping all other fields fixed. 
Isolate all terms in $S$ that depend on $\Phi_x$ and $\theta_x$.
{The process involves the following three steps:}
\paragraph{Neighbor sum.} 
For fixed $U$ and $\{\Phi_{y\neq x}\}$, define the gauge-covariant neighbor sum
\begin{align}
s_x = \sum_{\mu=1}^{d}\Bigl(U_{x,\mu}\Phi_{x+\hat{\mu}} + U_{x-\hat{\mu},\mu}^\dagger\Phi_{x-\hat{\mu}}\Bigr).
\end{align}
This collects contributions from both forward and backward links containing $x$.

\paragraph{Conditional for $\Phi_x$ given $\theta_x$.} 
The terms involving $\Phi_x$ are
\begin{align}
S_x = A|\Phi_x|^2 - \operatorname{Re}\bigl(\Phi_x^\dagger B\bigr) + \text{const},
\end{align}
with 
\begin{align}
A = d + \frac{1}{2\sigma^2},\qquad 
B = \frac{v e^{i\theta_x}}{\sigma^2} + s_x.
\end{align}
% Completing the square,
% \begin{align}
% A|\Phi_x|^2 - \operatorname{Re}(B^*\Phi_x) = A\Bigl|\Phi_x - \frac{B}{2A}\Bigr|^2 - \frac{|B|^2}{4A}.
% \end{align}
The conditional distribution is therefore a complex Gaussian
\begin{align}
{P(\Phi_x \mid \theta_x,s_x) = \frac{A}{\pi}\,
\exp\left(-A\Bigl|\Phi_x - \frac{B}{2A}\Bigr|^2\right)},
\end{align}
% with mean
% \begin{align}
% \mu_\Phi = \frac{B}{2A} = \frac{v e^{i\theta_x}/\sigma^2 + s_x}{2\bigl(d + \frac{1}{2\sigma^2}\bigr)}.
% \end{align}
% The variance in each real direction is $1/(2A)$. 
% This update is exact and efficient to sample.

\paragraph{Conditional for $\theta_x$ (marginalized over $\Phi_x$).} 
To update $\theta_x$, we integrate out $\Phi_x$ to get,
\begin{align}
P(\theta_x \mid s_x) 
\propto \exp\left(\frac{|B|^2}{4A}\right),
\end{align}
%Compute $|B|^2$:
where,
\begin{align}
|B|^2 = \frac{v^2}{\sigma^4} + |s_x|^2 + \frac{2v}{\sigma^2}\operatorname{Re}(s_x e^{-i\theta_x}).
\end{align}
Writing $s_x = |s_x|e^{i\psi_x}$, we have
\begin{align}
\frac{|B|^2}{4A} = \text{const} + \frac{v|s_x|}{2A\sigma^2}\cos(\theta_x - \psi_x).
\end{align}
Hence $\theta_x$ follows a von Mises distribution:
\begin{align}
{P(\theta_x \mid s_x) = \frac{1}{2\pi I_0(\kappa_x)}\,
\exp\Bigl(\kappa_x \cos(\theta_x - \psi_x)\Bigr)},
\end{align}
with concentration $\kappa_x = \frac{v|s_x|}{2A\sigma^2}$.
The mean direction is $\psi_x = \arg(s_x)$. 
Crucially, this conditional does not depend on the current $\Phi_x$, only on the neighbors $s_x$, which makes the Gibbs sampler particularly efficient.

\section{Finite size scaling}
\subsection{Finite-Size Scaling Analysis}

% We extract critical exponents from.
Observables such as the magnetization, susceptibility, and Binder cumulant exhibit finite-size scaling (FSS) behavior near the critical point, enabling the extraction of critical exponents that characterize the universality class.~\cite{Fisher1972, PrivmanFisher1984, Binder1981, JankeKenna2001, Privman1990, Cardy1988}. These observables are defined as follows. The magnetization 
\begin{equation}
  M = \langle |\Psi| \rangle, 
\end{equation} where \(\Psi = \sum_x \phi_x\) is the order parameter; the susceptibility 
\begin{equation}
 \chi = N \left( \langle \Psi^2 \rangle - \langle |\Psi| \rangle^2 \right), 
\end{equation}
with \(N = L^2\) the number of lattice sites, and  the Binder cumulant, defined as 
\begin{equation} 
U_4 = 1 - \frac{\langle \Psi^4 \rangle }{  3 \langle \Psi^2 \rangle^2}.
\end{equation}
 At the critical point, in the thermodynamic limit (\(L\to \infty\)), these observables obey the following scaling forms:
\begin{equation}
|M|(v_c, L)| \sim L^{-\beta/\nu},
\label{eq:magnetization_scaling}
\end{equation}
\begin{equation}
|\chi_{\text{max}}(L)| \sim L^{\gamma/\nu},
\label{eq:peak_scaling}
\end{equation}
\begin{equation}
\left|\frac{dU_4}{dv}\right|_{v_c} \sim L^{1/\nu},
\label{eq:binder_derivative_scaling}
\end{equation}
where \(v\) acts as the coupling parameter and \(v_c\) is the critical coupling. Additionally, the shift of the susceptibility peak position follows

\begin{equation}
|v_{\text{peak}}(L) - v_c| \sim L^{-1/\nu}.
\label{eq:shift_scaling}
\end{equation}
For the analysis presented here, we focus on the leading scaling behavior and ignore higher-order corrections, which are expected to be of order \(L^{-w}\), where \(w>0\) is the leading correction to the scaling exponent.

The critical coupling \(v_c\) is determined by the crossing of the Binder cumulant for different lattice sizes~\cite{Binder1981}; we define the crossing objective function
\begin{equation}
S(v) = \sum_{i<j} [U_4^{(i)}(v) - U_4^{(j)}(v)]^2,
\label{eq:crossing_objective}
\end{equation}
where \(U_4^{(i)}(v)\) are polynomial fits to the Binder cumulant data for each lattice size \(i\). The critical coupling \(v_c\) is then obtained as the value of \(v\) that minimizes \(S(v)\).

For each lattice size \(L\), the susceptibility maximum \(\chi_{\text{max}}\) and its location \(v_{\text{peak}}\) are determined by fitting the susceptibility data near the peak with a polynomial of degree \(d\):
\begin{equation}
\chi(v) = \sum_{n=0}^d a_n v^n.
\end{equation}
The peak position is then found from the condition \(d\chi/dv = 0\). The magnetization at \(v_c\) is obtained by interpolating \(|M(v)|\) at \(v_c\) using a local polynomial fit.

The critical exponents are extracted from log-log fits of the scaling relations in Eqs.~\eqref{eq:peak_scaling}\eqref{eq:magnetization_scaling}\eqref{eq:binder_derivative_scaling}\eqref{eq:shift_scaling}. For example, the exponent \(\gamma/\nu\) is obtained from the slope of \(\text{ln} \chi_{\text{max}}\) versus \(\ln L\). The error bars on all exponents are computed using a leave-one-chain-out jackknife procedure that propagates statistical uncertainties through the entire analysis pipeline, including the determination of \(v_c\).

To account for corrections to scaling, we perform the analysis using only lattice sizes that are sufficiently large to be in the asymptotic scaling regime.
\section{Numerical Experiments}

\begin{table}[htbp]
\centering
\caption{Comparison of MCMC estimates of exponents of the GMM-based distribution with Ising exponents in 2-dimensions.}~%\cite{Baxter1982}. }
\begin{tabular}{lcc}
\toprule
Exponent & GMM& Ising\\
\midrule
$\gamma/\nu$ & $1.75260(73)$& $1.75$ \\
$\beta/\nu$  & $0.1259(20)$& $0.125$ \\
$1/\nu$      & $0.995(15)$& $1.00$ \\
\bottomrule
\end{tabular}
\label{tab:exponents}
\end{table}

\begin{figure*}[t]  \label{fig:raw_observables}
    \centering
    \subfloat[][\label{fig:binder_crossing}]{%
        \includegraphics[width=0.45\textwidth]{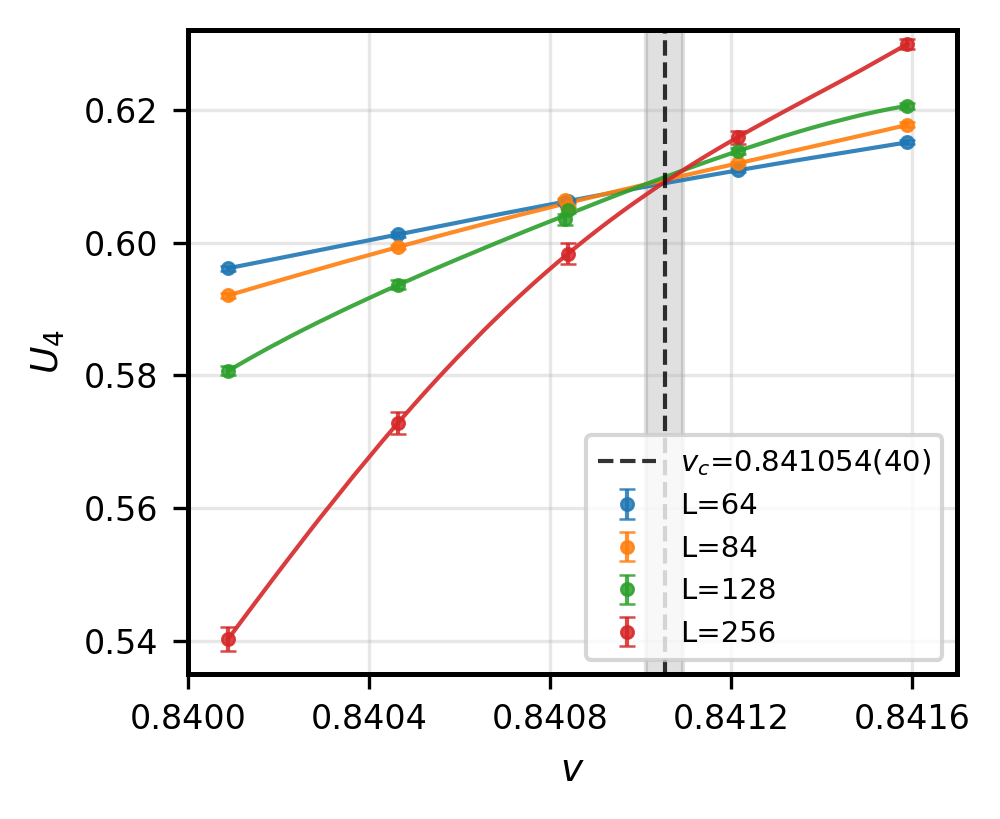}
    }
    \hfill
    \subfloat[][\label{fig:susceptibility}]{%
        \includegraphics[width=0.45\textwidth]{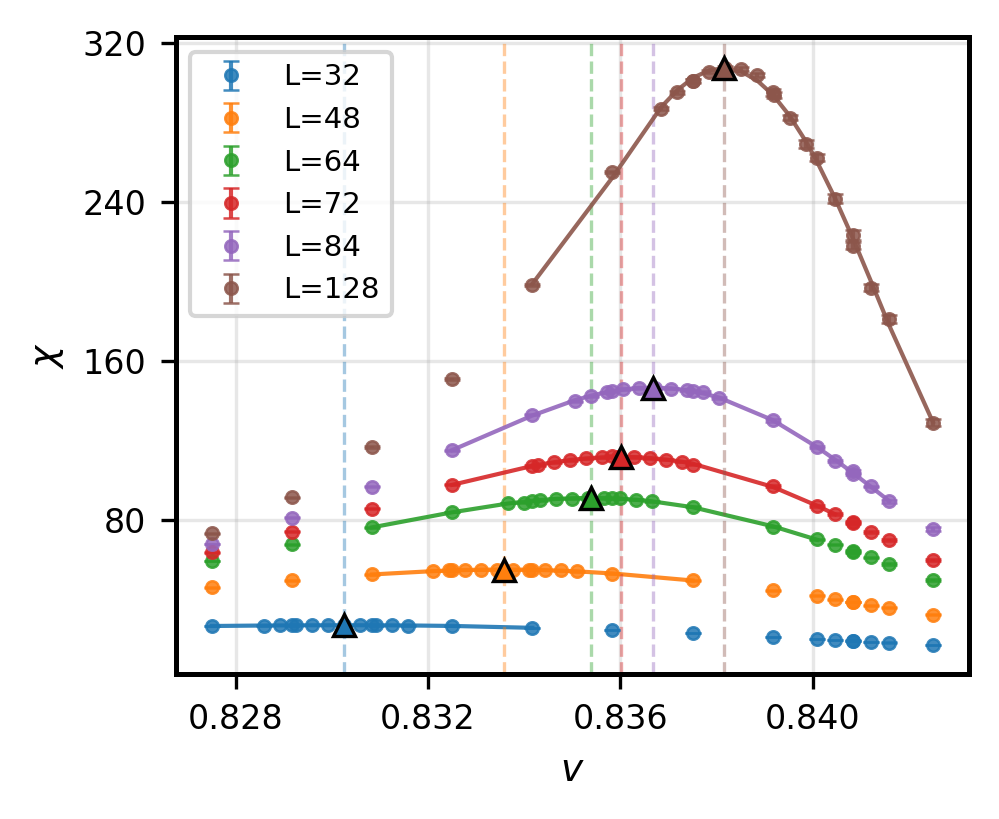}
    }
    % \hfill
    % \subfloat[][\label{fig:magnetization}]{%
    %     \includegraphics[width=0.32\textwidth]{Preliminary Figures/2D_Z2_magnetiation_at_peak_sus.png}
    % }
    \caption{(a) Binder crossing plot for lattice sizes 
                   $64^2,84^2, 128^2 $ and $256^2$, the vertical black dashed line shows the MCMC estimate of \(v_c\) .      
             (b) Susceptibility $\chi(v, L)$ for the lattice sizes $32^2, 48^2, 56^2,72^2,84^2$ and $128^2$. 
             The data points are fitted locally around each peak with polynomials 
             (solid lines) to extract precise peak positions $v_{\rm peak}(L)$ and 
             heights $\chi_{\max}(L)$. Triangle markers indicate the fitted peak locations. }

\end{figure*}

\begin{figure*}[t]
    \centering
    \subfloat[][\label{fig:chi_scaling}]{%
        \includegraphics[width=0.45\textwidth]{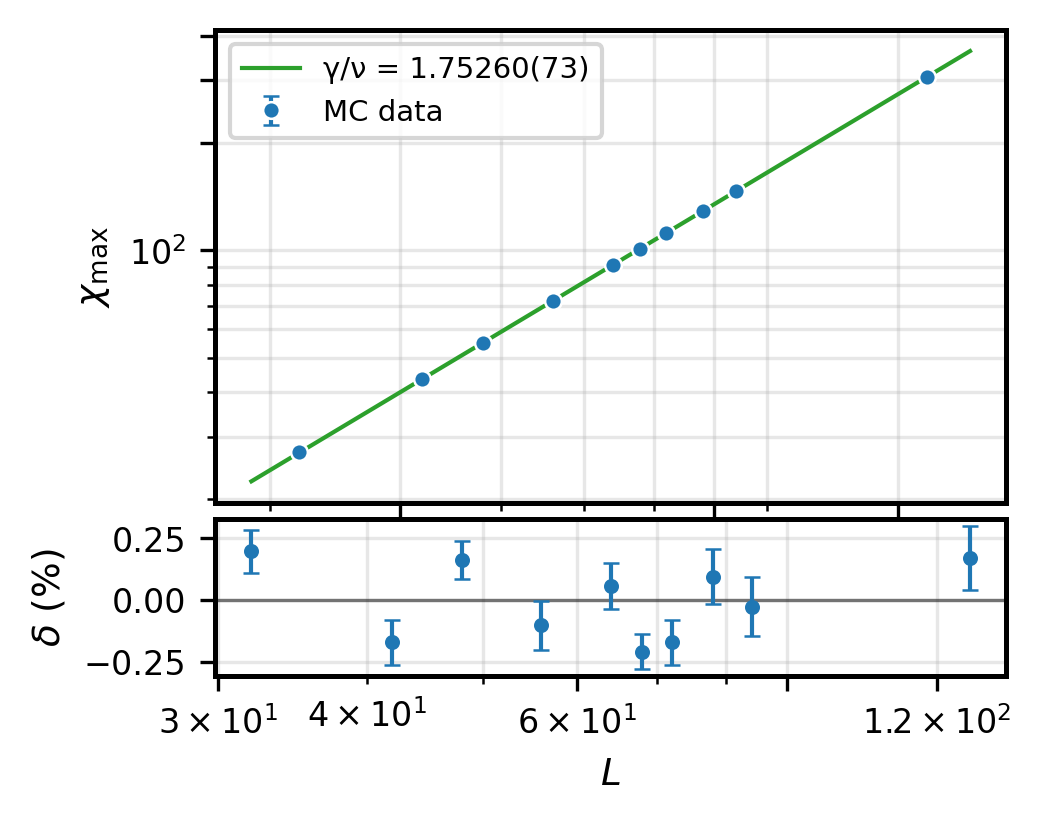}
    }
    % \hfill
    % \subfloat[][\label{fig:peak_shift}]{%
    %     \includegraphics[width=0.32\textwidth]{Final Figures/peak_shift_scaling.png}
    % }
    \hfill
    \subfloat[][\label{fig:mag_scaling}]{%
        \includegraphics[width=0.45\textwidth]{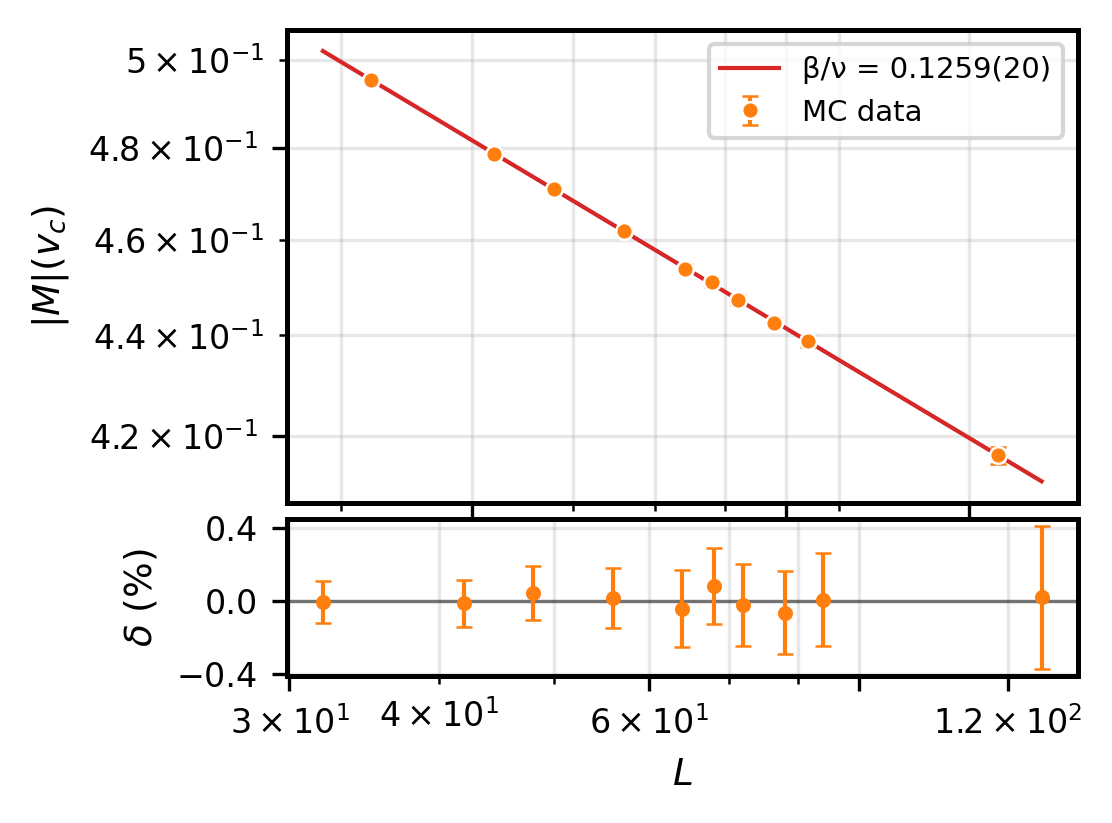}
    }
    \caption{(a) Finite-size scaling of the susceptibility peak $\chi_{\max}(L)$. 
             The fitted exponent is $\gamma/\nu = 1.75260(73)$, in agreement with 
             the exact 2D Ising value $\gamma/\nu = 7/4 = 1.75$. The lower 
             panel shows the percentage residuals.
             (b) Finite-size scaling of the magnetization at criticality 
             $|m(v_c)|$. The fitted exponent $\beta/\nu = 0.1259(20)$ agrees 
             with the exact value $\beta/\nu = 1/8 = 0.125$. The lower panel 
             displays the percentage residuals.}
    \label{fig:beta_gamma_scaling}
\end{figure*}

We have sampled the action in Eq.~\eqref{eq:Z2_GMM_action} on two-dimensional square lattices with periodic boundary conditions, 
and verified that it belongs to the two-dimensional Ising universality class by measuring critical exponents using 
Markov chain Monte Carlo (MCMC) methods. We have also verified the Ising limit of the joint action \(S[\phi,z]\)
in Eq.~\eqref{eq:Z2_GMM_joint_action}, which is given by Eq.~\eqref{eq:Ising_limit}.

Several sampling strategies have been explored in this work. In addition to the heat-bath updates described in Sec.~\ref{subsec:Heat_Bath_Updates} and the local Hamiltonian dynamics of Sec.~\ref{sec:local_hamiltonian_dynamics}, the auxiliary-field representation in Eq.~\ref{eq:Z2_GMM_joint_action} also enables global Hamiltonian updates. In this approach, the global Hamilton's equations are coupled in position space but decouple in Fourier space, allowing for efficient updates that can be applied in Fourier space and then transformed back to real space. In practice, we find that Hamiltonian methods perform well for larger values of \(\sigma\), but suffer from slow thermalization for smaller \(\sigma\). The heat-bath method, on the other hand, works reliably across the entire range of \(\sigma\) values. For this reason, the results presented in this work are obtained using the heat-bath update rule.
For each lattice size, we generated 20 independent Markov chains with different random 
seeds. The first \(4 \times 10^4\) full sweeps of each chain were discarded as 
thermalization (burn-in) and were not included in the recorded data. Subsequently, 
each chain was run for \(10^7\) full sweeps, from which configurations were recorded 
for analysis. Each full sweep comprised two half-sweeps in the odd-even checkerboard 
pattern, updating all sites of one parity followed by all sites of the opposite parity. 
This checkerboard decomposition allows for efficient parallelization, as sites of the 
same parity are non-interacting and can be updated simultaneously. To reduce 
autocorrelation among the recorded samples, we applied a thinning procedure by 
recording configurations every \(10^3\) full sweeps.
All error bars in further analysis are computed using a blocked jackknife procedure~(\cite{Kunsch1989, Young1996}), where each jackknife block corresponds to one of the 20 independent Markov chains. 

\begin{figure*}[t]
    \centering
    \subfloat[][\label{fig:shift_scaling}]{%
        \includegraphics[width=0.45\textwidth]{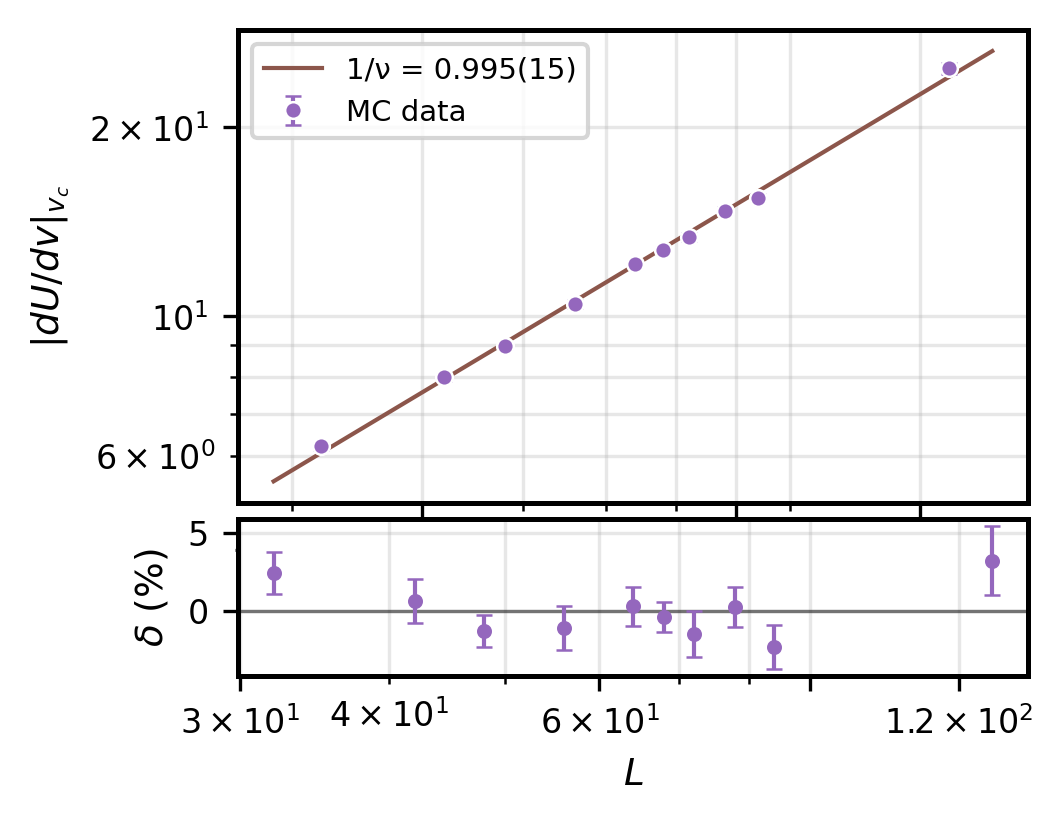}
    }
    \hfill
    \subfloat[][\label{fig:derivative_scaling}]{%
        \includegraphics[width=0.45\textwidth]{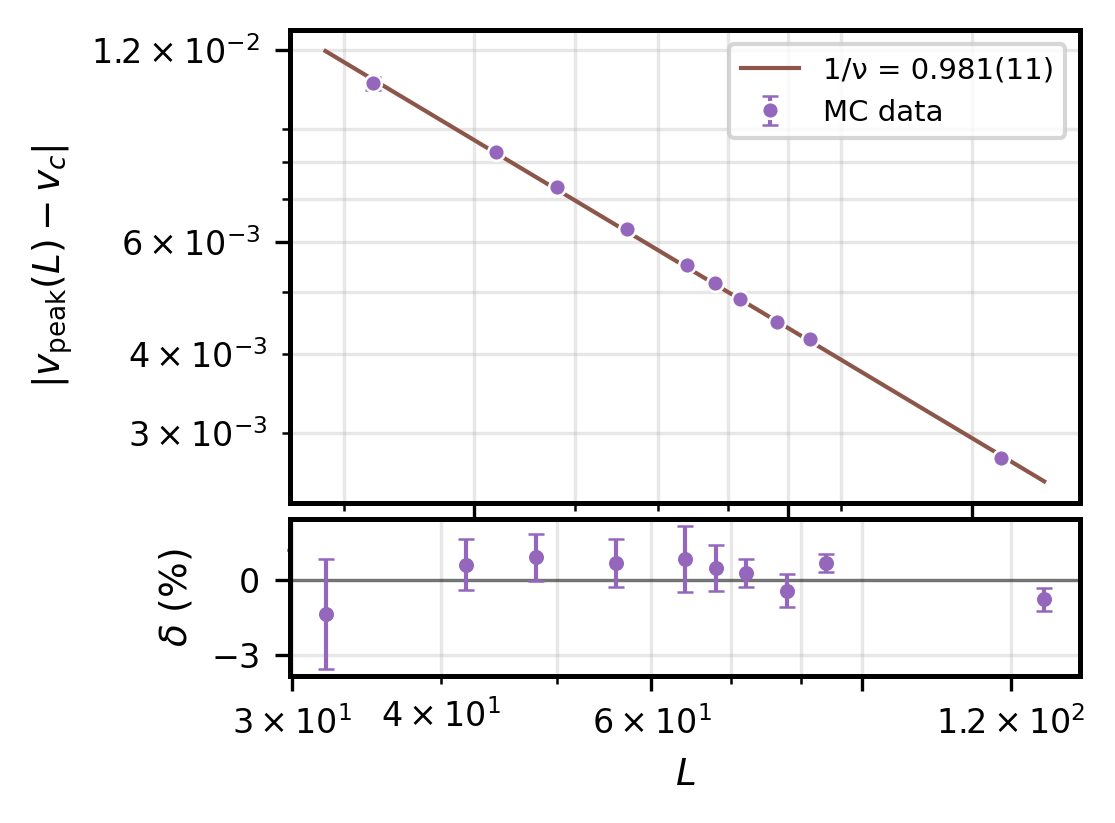}
    }
    % \hfill
    % \subfloat[][\label{fig:magnetization}]{%
    %     \includegraphics[width=0.32\textwidth]{Preliminary Figures/2D_Z2_magnetiation_at_peak_sus.png}
    % }
    \caption{(a) Finite-size scaling of the  derivative of \(U_4\) w.r.t. \(v\) at \(v_c\), 
             $\left|\frac{dU_4}{dv}\right|_{v_c}$. The fitted exponent $1/\nu = 0.995(15)$, consistent with exact value $\nu = 1$. The lower panel 
             displays the percentage residuals.
    (b) Finite-size scaling of the pseudocritical point shift 
             $|v_{\rm peak}(L) - v_c|$. The fitted exponent $1/\nu = 0.981(11)$, somewhat different from the exact value. The lower panel 
             displays the percentage residuals.}
    \label{fig:nu_scaling}
\end{figure*}

\subsection{Verifying 2D Ising exponents}

We now perform a detailed FSS analysis to extract the critical 
exponents corresponding to the action given by Eq.~\eqref{eq:Z2_GMM_action} in two dimensions:$\gamma/\nu$, $\beta/\nu$, 
and $1/\nu$,
% ~(\ref{eq:peak_scaling},\ref{eq:magnetization_scaling},\ref{eq:binder_derivative_scaling},\ref{eq:shift_scaling})
presented in Table~\ref{tab:exponents}. We have performed FSS for a fixed $\sigma^2=0.2$, this choice is motivated to reduce finite-size effects, for large \(\sigma^2\) the potential in Eq.~\eqref{eq:Z2_GMM_potential} is broad, suppressing the self-interaction of the scalar field \(\phi\), resulting in a very large correlation length at criticality, and we need larger lattices in comparison with the correlation length.

Figure~\ref{fig:binder_crossing} shows the Binder cumulant crossing analysis. To extract $v_c$ precisely, we fit each $U_L(v)$ 
curve with a polynomial of degree 4 within the crossing region 
$v \in [0.8400, 0.8416]$ and determine the intersection by minimizing the objective function in Eq.~\eqref{eq:crossing_objective} using the fitted curve values. Figure~\ref{fig:susceptibility} shows the susceptibility data. 
The susceptibility peaks sharpen and shift toward $v_c$ with increasing lattice size ($L$).

To compute the position of \(\chi_{max}\), we fit each \(\chi\) curve with a polynomial of degree 4. Similarly, for estimating \(|M|_{v_c}\) for each L, we fit each \(|M|\) curve with a polynomial of degree 4 and to estimate \(\left|\frac{dU_4}{dv}\right|_{v_c}\), we fit each \(U_4\) curve with a polynomial of degree 6 in region \(v\in [0.8200, 0.8420]\).
% while the magnetization curves become steeper near $v_c$, reflecting the 
% development of a non-zero order parameter in the thermodynamic limit. 

Figures~\ref{fig:chi_scaling} and~\ref{fig:mag_scaling} show the scaling laws for \(\chi_{max}\) and \(|M|_{v_c}\) given by Eqs.~\eqref{eq:peak_scaling} and~\eqref{eq:magnetization_scaling}, respectively. 
Figures~\ref{fig:shift_scaling} and~\ref{fig:derivative_scaling} show the scaling law for the  derivative of \(U_4\) w.r.t. \(v\) at \(v_c\) and  peak shifting of \(\chi\) given by Eqs.~\eqref{eq:binder_derivative_scaling} and~\eqref{eq:shift_scaling} respectively, both depend on the exponent \(\nu\) providing two different methods to estimate \(\nu\). Using the exponents from Table~\ref{tab:exponents}, the hyperscaling relation 
$2\beta/\nu + \gamma/\nu = d$, with $d=2$, gives 
$2(0.1259(20)) + 1.75260(73) = 2.0044(43)$, 
while the Fisher relation $\eta = 2 - \gamma/\nu$ yields 
$\eta = 0.24740(73)$, both consistent with the exact 2D Ising values 
$2$ and $0.25$, respectively.

\subsection{The limit $\sigma \to 0$}

To verify that our Gaussian mixture model (GMM) heat-bath sampler correctly 
reproduces the 2D Ising model in the limit 
$\sigma^2 \to 0$, we perform binder-crossing analysis at $\sigma^2=0.0001$, with lattice sizes:  \(48^2, 64^2, 128^2\). 

\begin{figure}[htbp]
    \centering
    \includegraphics[width=0.9\columnwidth]{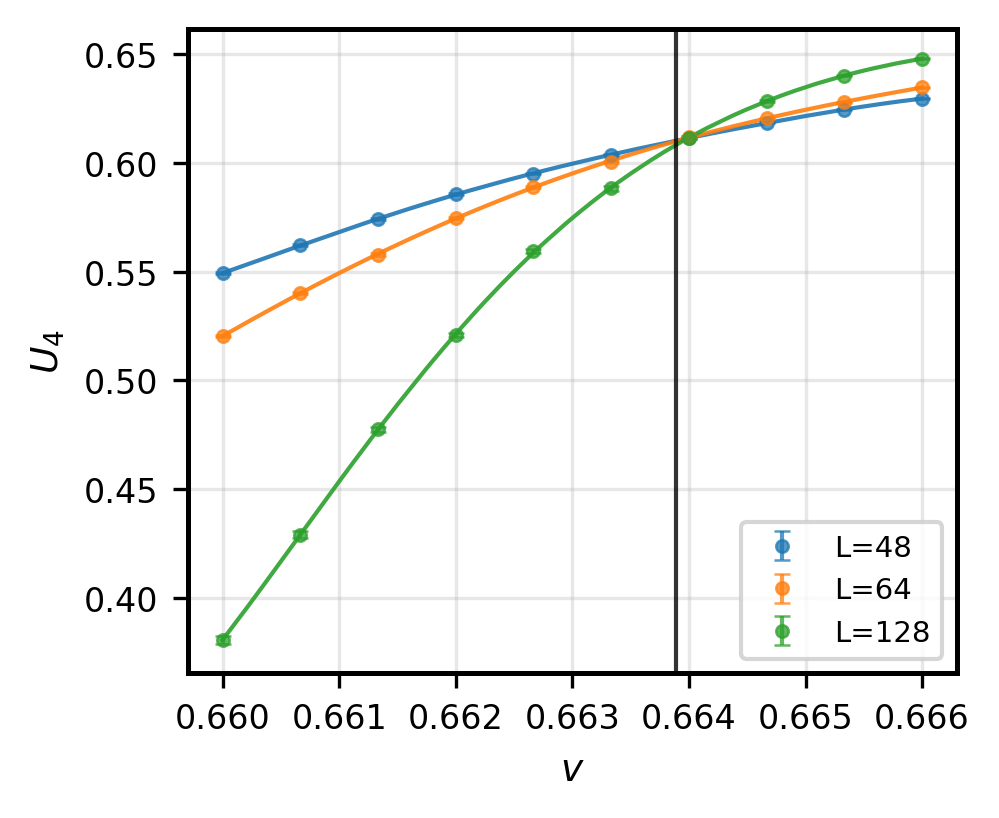}
    \caption{Binder cumulant $U_L$ as a function of the coupling parameter $v$ 
             for lattice sizes $ 48^2, 64^2$ and $128^2$. The curves are fitted with 
             polynomials (solid lines) within the crossing region. The vertical 
             solid black line shows the exact Onsager value,
             $v_c^{\rm exact}  \approx 0.66389$. While the MCMC estimate at \(\sigma^2=0.0001\) is, $v_c = 0.66399(4)$.}
    \label{fig:gmm_ising_limit}
\end{figure}

Figure~\ref{fig:gmm_ising_limit} demonstrates that as $\sigma$ becomes 
sufficiently small, the critical point converges to the known 
2D Ising values. This establishes the correct limiting behavior of the 
GMM potential.

\section{Conclusion}
Studying the criticality of scalar field theory is important as it provides a framework to study the continuous phase transition in both condensed matter physics and quantum field theory. It is also important to find the exact critical point at which the theory becomes conformally invariant. Studying criticality in a lattice field theory is always a challenging problem due to the diverging correlation length near the critical point. 
We have proposed a class of GMM-based scalar potentials for which heat-bath sampling is as easy as for spin models with the corresponding internal symmetry group; furthermore, when coupled to gauge field sampling, the scalar field is still as cheap as a spin model. We have also shown that our GMM-based two-dimensional \(Z_2\) symmetric scalar theory belongs to the two-dimensional Ising universality class by numerically computing the critical exponents. One needs to take sufficiently large lattices to ensure that the correlation length does not grow larger than the lattice size ($L$). Continuous unconstrained scalar variables do carry more information than constrained spin variables, as they do have radial fluctuations, allowing these GMM-based scalar potentials to be useful for studying the important physics of spontaneous symmetry breaking, Higgs phenomena, and Gradient Flow studies numerically on lattice-discretised scalar fields, at lower computational cost than polynomial potentials. It will be interesting to extend the framework to study the system with disorder and defects, also for field theories including coupling with gauge fields in higher dimensions.

% \clearpage
% \FloatBarrier 

\end{document}